\documentclass[conference,a4paper]{APSIPA2019}
\usepackage{multirow}
\usepackage[dvips]{graphicx}
\usepackage{amsmath}
\usepackage[psamsfonts]{amssymb}
\usepackage{amsxtra}
\usepackage{threeparttable}

\renewcommand{\baselinestretch}{0.97}

\begin{document}

\title{Real-time and interactive tools for vocal training based on
an analytic signal\\ with a cosine series envelope}

\author{%
\authorblockN{%
Hideki Kawahara\authorrefmark{1},
Ken-Ichi Sakakibara\authorrefmark{2},
Eri Haneishi\authorrefmark{3} and
Kaori Hagiwara\authorrefmark{3}
}
\authorblockA{%
\authorrefmark{1}
Wakayama University, Wakayama, Japan \\
E-mail: kawahara@wakayama-u.ac.jp}
\authorblockA{%
\authorrefmark{2}
Health Science University of Hokkaido, Sapporo, Japan\\
E-mail: quesokis@gmail.com}
\authorblockA{%
\authorrefmark{3}
Showa University of Music, Kawasaki, Japan\\
E-mail: haneishi@tosei-showa-music.ac.jp, k-hagiwara@tosei-showa-music.ac.jp}
}

\maketitle
\thispagestyle{empty}

\begin{abstract}
We introduce real-time and interactive tools for assisting vocal training.
In this presentation, we demonstrate mainly a tool based on real-time visualizer of fundamental frequency candidates
to provide information-rich feedback to learners.
The visualizer uses an efficient algorithm using analytic signals for deriving phase-based attributes.
We start using these tools in vocal training for assisting learners to acquire
the awareness of appropriate vocalization.
The first author made the MATLAB implementation of the tools open-source.
The code and associated video materials are accessible in the first author's GitHub repository.
\end{abstract}

\section{Introduction}


Adequate vocalization, such as speaking and singing,  
plays an essential role in maintaining a  good quality of daily human life.    
For old adults,
adequate vocalization helps their quality of life by promoting social communication activities and, 
in some cases, even reducing
the risk of aspiration pneumonia.   
For vocal professionals,  such as singers, teachers, and music therapists,
proper vocalization technique is crucial for preserving their professional
lives\cite{haneishi2001effects}.   
To maintain proper vocalization skills, we introduce tools for assisting vocal training.  
Real-time tools provide visual feedback of voicing attributes. 
The supporting tools enable detailed analysis and study the effects of articulatory changes.    
The tools use an extractor of $f_\mathrm{o}$\footnote{Using ``$f_\mathrm{o}$'' instead of ``F0'' for representing
fundamental frequency is based on a recommendation\cite{titze2015jasaforum}}
(fundamental frequency) candidates in real-time and audio sampling rate
using analytic signals with a six-term cosine series\cite{kawahara2017interspeechGS}. It
also uses other interactive tools for visualization of auditory
information and speech production processes\cite{Kawahara2016isst}.
The body part of this article focuses on the introduction
and use of the tools.  
We placed technical details in the appendix for making the body text less confusing.

\section{Real-time and interactive tools}
This section mainly focuses on the tool based on a real-time extractor of $f_\mathrm{o}$ candidates.
In a training session, the instructor uses this tool interactively to demonstrate learners how to control voicing
by visualizing their voices in real-time and playback the reference samples and the learners' voices.

The tool has the following functions; a)$f_\mathrm{o}$ candidates' trajectory visualization with the salience of periodicity, b)musical note display, c)waveform display (long and short duration), d)sound pressure level, e)spectrum, f)recording with calibration, and g)playback of working and reference recordings.

\begin{figure}[tbp]
\begin{center}
\includegraphics[width=\hsize]{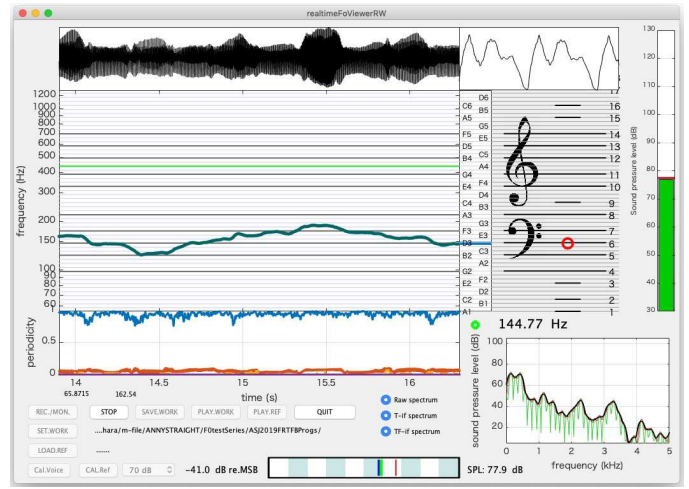}
\caption{GUI of the tool with a real-time $f_\mathrm{o}$ candidates extractor.}
\label{realtimeVocalInspectorGUImain}
\end{center}
\end{figure}
Figure~\ref{realtimeVocalInspectorGUImain} shows a snapshot of the GUI of the tool.%
\footnote{The first author's GitHub repository\cite{kawaharaGit} has the latest version of the accompanying video,
which introduces how to use this tool.}
The left three panels are continuously scrolling to the left to show trajectories of the waveform, $f_\mathrm{o}$ candidates, and the periodicity salience (from top to bottom). 
The top right panel shows the stabilized waveform of several cycles of the fundamental period. 
The middle right panel shows staves with a treble clef and a bass clef. 
We introduced this panel to help learners to be aware of voice pitch because sometimes learner has difficulty in finding what ``high'' or ``low'' pitch is. 
The red circle represents the corresponding $f_\mathrm{o}$ position.
Each stave has a number at the right side. 
The instructor can use the red circle and the number to advise learners. 
Just below the musical panel, a green circle and a number representing the frequency of the best $f_\mathrm{o}$ candidate appear when the signal periodicity is salient. 
The small panel between the scrolling display and the musical display shows corresponding note names. 
Thin horizontal lines span these three panels. 
They represent the chromatic scale.

The rightmost bar graph shows the calibrated sound pressure level 
at the reference position (30~cm in front of the lip center\cite{Rita2018ajsp}) using C-weighting\cite{IEC61672}.
The green bar shows the fast response, and the horizontal red line shows the slow response.

The bottom right panel shows spectral information using three lines.
The three lines are a)power spectrum (a green line), b)temporally interference-free spectrum (a red line), and
c)spectrum without temporal and frequency interferences (a bold black line)\cite{kawahara2018apsipa}.
The vertical axis represents the sound pressure level when calibrated.

The bottom center bar panel is a level indicator in dB scale spanning from -100~dB to 0~dB, where
0~dB corresponds to the MSB (Most Significant Bit, the maximum input level).
This indicator has three cursors. 
The red cursor represents the peak value, and
the blue cursor represents the RMS value.
Finally, the green cursor represents the temporally smoothed RMS value.
The user can calibrate this tool by
using any stable 
test signals (including the learner's sustained vowel) and a sound level meter together with this level indicator.
Please refer\cite{svec2010ajsp,Rita2018ajsp} for the recommended procedure and microphone settings.

This tool has ten push buttons and a popup menu. 
They have the following functions.
``REC. START'' button restarts the tool.
The ``SAVE.WORK'' button writes the contents of the input buffer to a file providing a unique name.
The saved file is a WAVE format with 44,100~Hz sampling frequency and 24-bit resolution.
The ``STOP'' button stops the continuous monitoring of the input.
``PLAY.WORK'' button outputs the signal in the audio input buffer.
``PLAY.REF'' button outputs the signal in the reference buffer.
The ``QUIT'' button terminates the tool.
``SET.WORK'' button assigns the directory for writing work files.
``LOAD.REF'' button reads an audio file as the reference.
``Cal.Voice'' and ``CAL.Ref'' buttons calibrate the microphone input level.
Only adequate buttons are active and highlighted. 

The popup menu showing ``70 dB'' selects the calibration sound pressure level.
Our GitHub repository\cite{kawaharaGit} has links to videos showing how to calibrate this tool
and several operation examples.

\subsection{Supporting interactive tools}
In addition to this tool, we developed two supporting tools and applied
previously developed tools\cite{Kawahara2016isst}.
Developed tools are a real-time phase-attributes visualizer and
a detailed interactive inspector.

\subsubsection{Real-time phase-attributes visualizer}
\begin{figure}[tbp]
\begin{center}
\vspace{2mm}
{~ }\includegraphics[width=0.9\hsize]{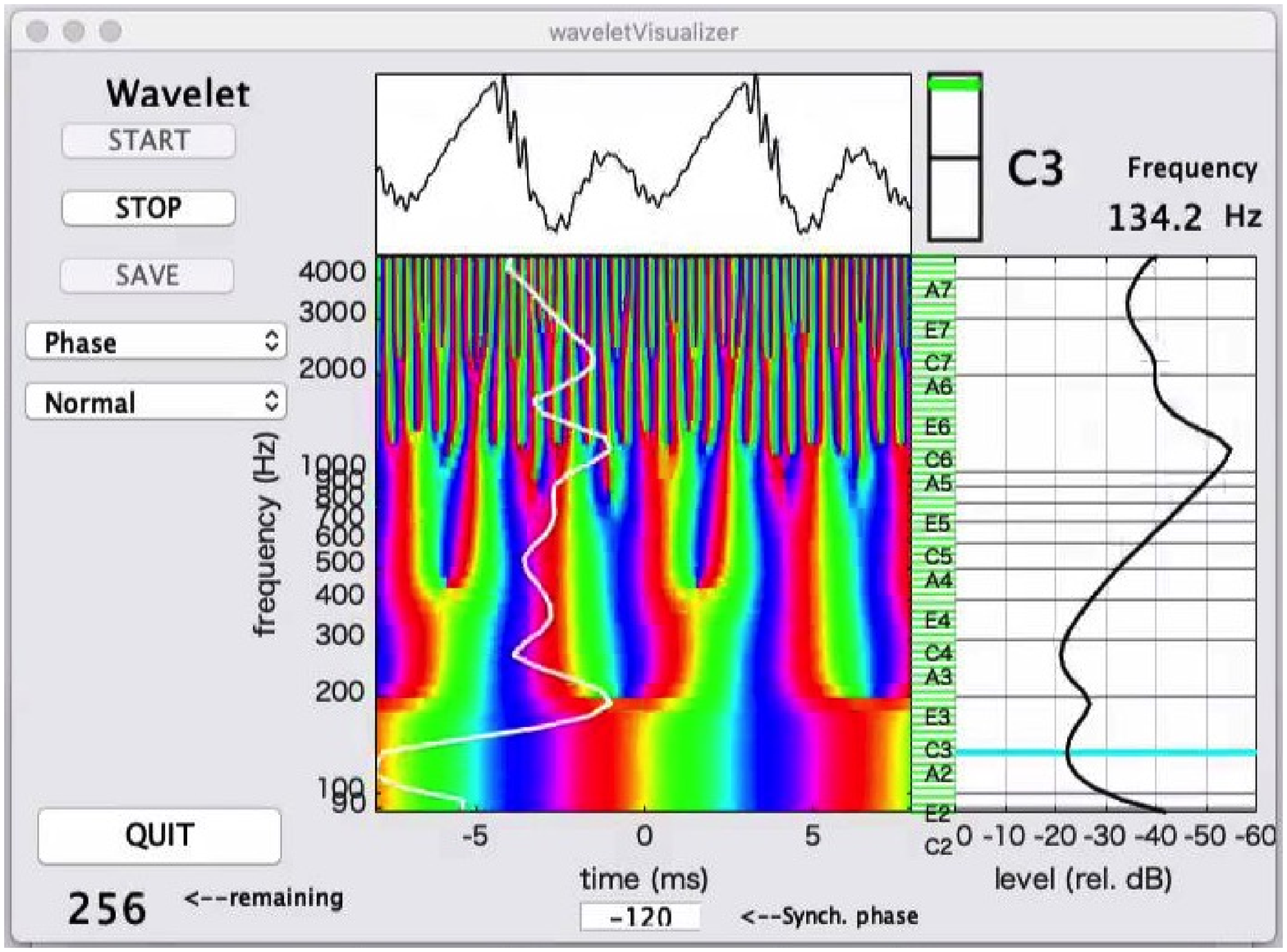}\\ 
\vspace{2mm}
{~ }\includegraphics[width=0.9\hsize]{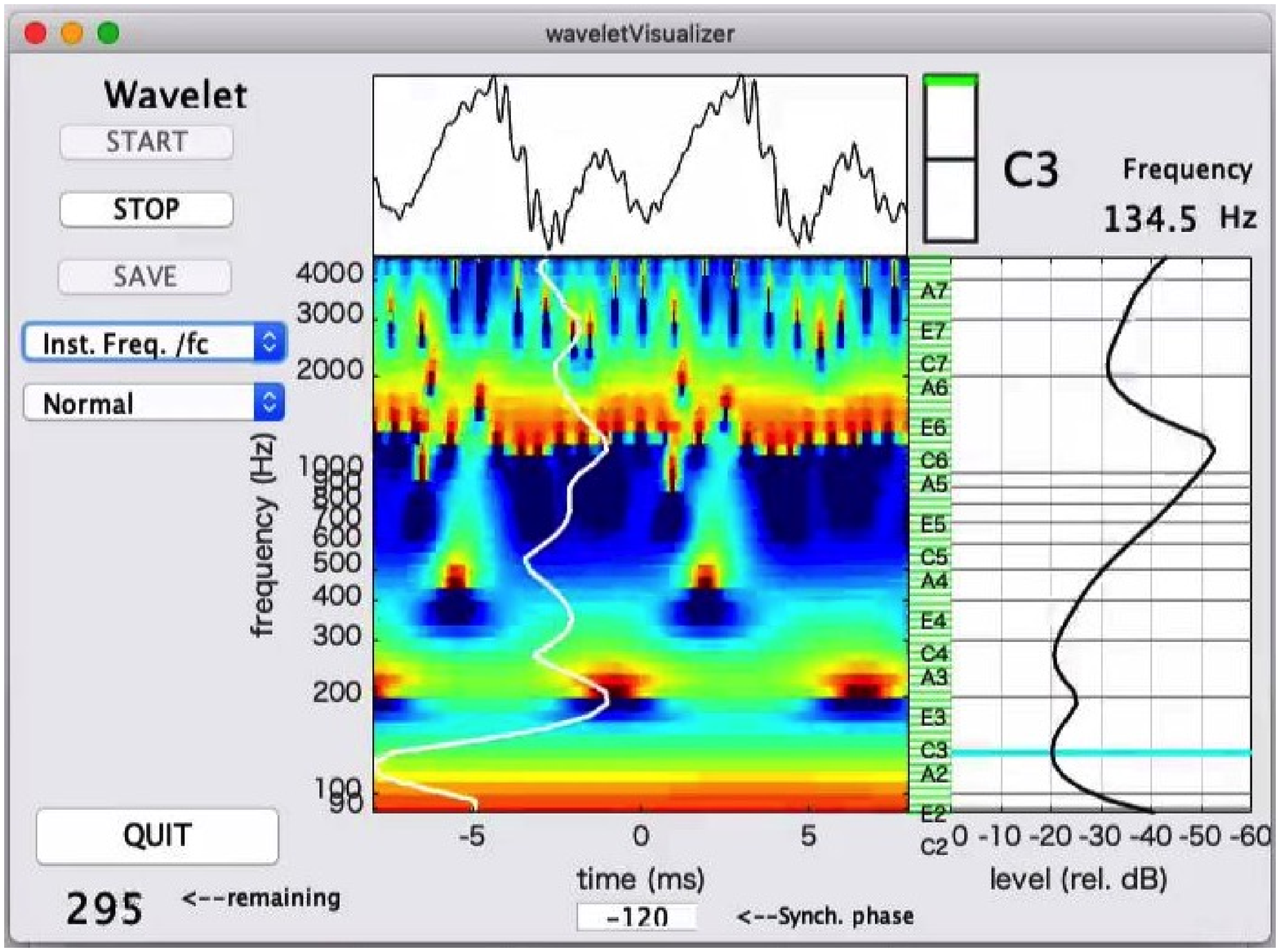}\\ 
\vspace{2mm}
{~ }\includegraphics[width=0.9\hsize]{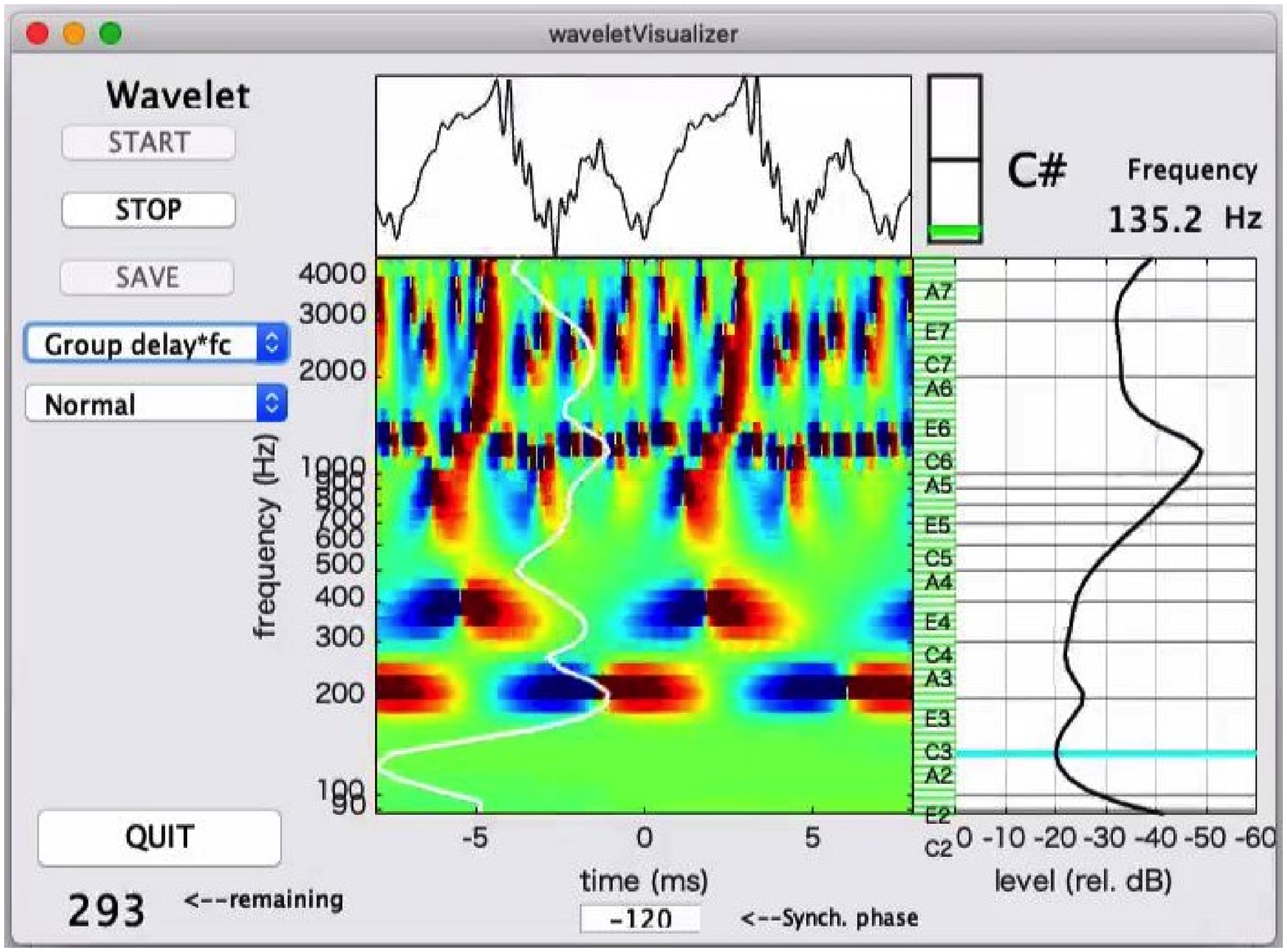}
\caption{Snapshots of a real-time phase-attributes visualizer.}
\label{wvltPhase}
\end{center}
\end{figure}
Figure~\ref{wvltPhase} shows example visualizations of three phase-related attributes.
From top to bottom, the panels show the phase map, the normalized instantaneous frequency map, and
the normalized group delay map.

Descriptions of the components in each panel are as follows.
This description is for the top panel.
The top left panel shows the pseudo color image of the phase of the analysis results in the middle.
The line plot above the image shows the time-aligned input waveform based on
the phase of the fundamental component. 
The cyan line in the right plot indicates the fundamental component.
The rightmost line plot shows the RMS (root mean squared) average of each output.
The top right corner displays the frequency of the best $f_\mathrm{o}$ candidate and the closest musical note name with 
a tuning display (The black center line shows the target note for the green line pitch monitor).
The main image of this panel shows the pseudocolor representation of the phase value itself.
We use the colormap which yields cyclic impression because phase has a circular topology.

The second panel shows the relative output instantaneous frequency normalized by the center frequency.
The relative instantaneous frequency display has a rainbow-colored horizontal bar 
which corresponds to the fundamental component.

The bottom panel shows the relative output group delay normalized by the nominal period of each filter response.
The relative group delay displays have (gently bending) rainbow-colored vertical bars which correspond to the
GCIs (Glottal Closure Instances).

\begin{figure}[tbp]
\begin{center}
{~ }\includegraphics[width=0.9\hsize]{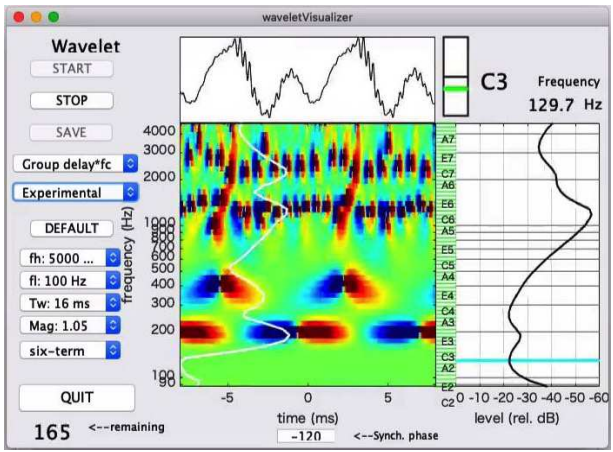}
\caption{Snapshot of the ``Experimental'' mode of a real-time phase-attributes visualizer.}
\label{wvltGourDelayEx}
\end{center}
\end{figure}
Figure~\ref{wvltGourDelayEx} shows the ``Experimental'' mode, which allows an extended parameter setting.

\subsubsection{Detailed interactive inspector}

\begin{figure}[tbp]
\begin{center}
\includegraphics[width=\hsize]{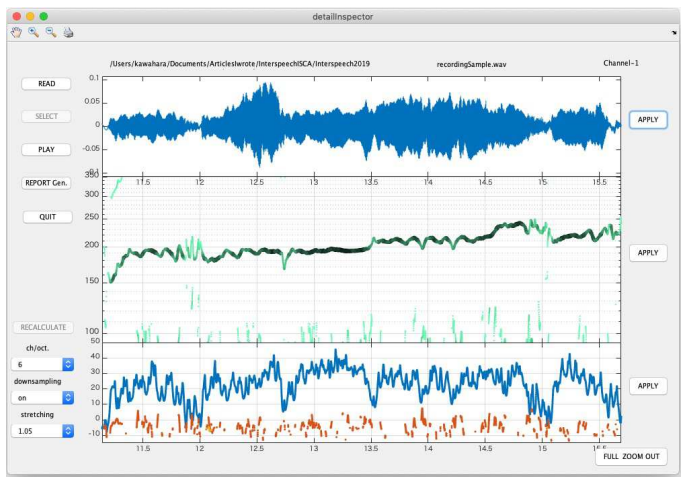}
\caption{GUI of an interactive detailed inspector.}
\label{detailExpandMagAVCTK}
\end{center}
\end{figure}
Figure~\ref{detailExpandMagAVCTK} shows a snapshot of the detailed inspector.
The analyzed file is the contents of the real-time tool of the session shown in the accompanying video.
The top panel shows the waveform for about 4.5 seconds.
The middle panel shows the $f_\mathrm{o}$ candidates trajectories.
The thickness and the darkness of each trajectory represent salience of the periodicity.
The bottom panel represents periodicity salience.  
It uses the estimated SNR for the salience. 
The user can interactively check details by ``zoom in,'' ``zoom out,'' and ``panning'' in each panel.

\subsubsection{SparkNG: previous tools}

\begin{figure}[tbp]
\begin{flushright}
\hspace{3mm}
\includegraphics[width=\hsize]{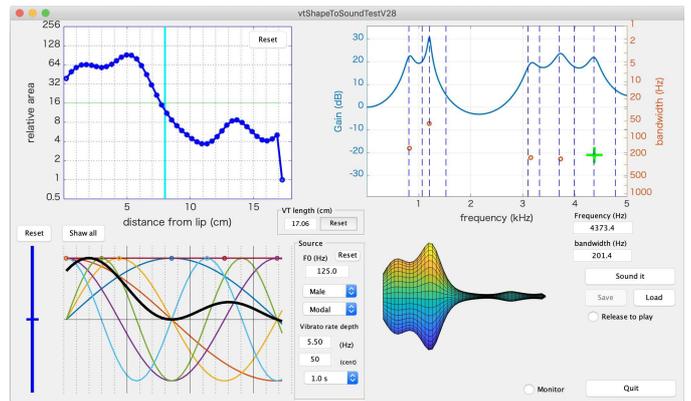}
\caption{GUI of a voice production simulator in
the previous tools\cite{Kawahara2016isst}.}
\label{vtshapeSimulatorGUI}
\end{flushright}
\end{figure}
Figure~\ref{vtshapeSimulatorGUI} shows the snapshot of the GUI of a voice production simulator in our previous tools\cite{Kawahara2016isst}.
The tool provides interactive manipulation of voicing attributes and provide results by synthesizing voiced sounds.
The attributes are the vocal tract area function, the first five formant frequencies and bandwidths,
vocal tract length, and the shape parameters of the glottal source waveform.
The results of parameter modifications directly change the synthesized sounds and
enabling users to acquire intricate relations between various voicing attributes.

\section{Conclusions}

We introduced real-time and interactive tools for assisting vocal training.
The tools are implemented using MATLAB and open-sourced.
The introduced tools and the other assistive tools, SparkNG are available in
the first author's GitHub repository\cite{kawaharaGit}.

\section{Acknowledgements}

This research was supported by KAKENHI (Grant in Aid for Scientific Research by JSPS) 16H01734, 15H03207, 18K00147, and 19K21618.

\renewcommand{\textheight}{235mm}

\bibliographystyle{IEEEtran}

\bibliography{APSIPA19RTFB.bib}

\appendix

\subsection{Real-time $f_\mathrm{o}$ candidates extractor}
This appendix briefly describes the
procedures used to implement the proposed tools.

\subsubsection{Simplified implementation of phase-related attributes}
Flanagan introduced an equation to calculate the instantaneous frequency,
which does not rely on phase unwrapping\cite{flanagan66jasa}.
Recent advances in multimedia processing made
CPU have efficient instructions for calculating related functions (for example\cite{intelVMperformance2018}).

These modern CPUs calculates the following discrete implementation
of instantaneous frequency $\omega_i[n]$ and group delay $\tau_g[k]$ efficiently.
\begin{align}\label{eq:directif}
\omega_i[n] & = \angle\left[\frac{x[n+1]}{x[n]} f_s\right] , \\
\tau_g[k] & = -\frac{1}{\Delta \omega} \angle\left[\frac{X[k+1]}{X[k]} \right] ,
\end{align}
where $x[n]$ represents an analytic discrete-time signal,
and $k$ represents the discrete frequency.

\subsubsection{Analytic signal with the six-term cosine series envelope}
Filtering an input signal using a filter having
an analytic signal impulse response yields
an analytic signal output.
Multiplying a time windowing function and a carrier complex exponential function makes such analytic signal.
We found that commonly used time windowing functions\cite{harris1978ieee,kaiser1980use,nattall1981ieee} are not
relevant for calculating phase-related attributes\cite{kawahara2017interspeechGS}.
The following function $w_e(t; f_c, c_\mathrm{mag})$ provides practically the best results.
\begin{align}
 w_e(t; f_c, c_\mathrm{mag}) & = \sum_{k=0}^{K} a_k \cos\left(\frac{2 \pi k f_c t}{K c_\mathrm{mag}} \right) ,
\end{align}
where $c_\mathrm{mag}$ represents a tuning parameter
which determines the relation between the bandwidth and the carrier frequency $f_c$.
We used the following cosine series with $K=5$.
The optimized coefficients are the following.
\begin{align}
\{a_k \}_{k=0}^5 & =  \left\{0.2624710164, 0.4265335164,  0.2250165621, \right. \nonumber \\
 &  \!\!\!\!\!\! \left. 0.0726831633,  0.0125124215, 0.0007833203\right\} \label{eq:sixTerm}
\end{align}
The following equation provides the desired impulse response.
\begin{align}
w(t) & = w_e(t; f_c, c_\mathrm{mag}) \exp\left( j 2 \pi f_c t \right)  \ \
\mbox{where} \ \ j = \sqrt{-1} ,
\end{align}

\subsubsection{SNR estimation}
When the dominant component is a stable sinusoid located in the pass-band of a filter,
the instantaneous frequency of the filter output is temporally constant.
The group delay calculated using adjacent channel outputs
is also temporally constant.
Mixing temporal variations of the instantaneous frequency and
the group delay provides the estimate of the SNR of the filter output.

\begin{figure}[tbp]
\begin{center}
\includegraphics[width=0.8\hsize]{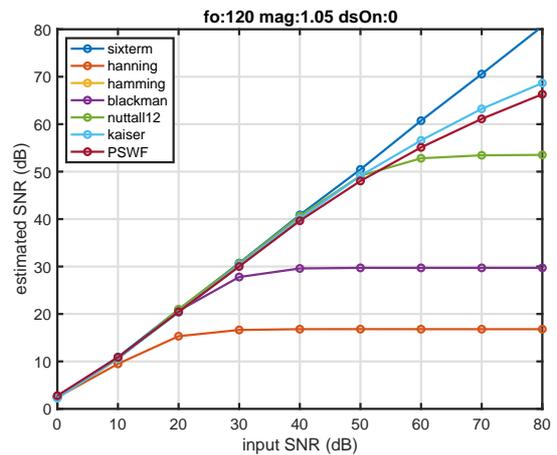}
\caption{Estimated SNR for each envelope functions.
The stretching factor is $c_\mathrm{mag} = 1.05$.}
\label{winSNRcalibration0fo120Hz}
\end{center}
\vspace{-3mm}
\end{figure}
Figure~\ref{winSNRcalibration0fo120Hz} shows the relation
between the given SNR and the estimated SNR using commonly used windowing functions
and the proposed six-term cosine series.
The proposed envelope yields
accurate estimates from 10~dB to 80~dB SNR.
This SNR estimation executes about 300 times faster than real-time for
a single filter output for 44,100~Hz sampled signals.
This speed test used MATLAB implementation on MacBook Pro
(2019 13" 2.7~GHz Intel Corei7, with 16~GB memory).

\subsubsection{Extractor of $f_\mathrm{o}$ candidates}
We placed six SNR estimators for each octave covering from 80~Hz to 5~kHz.
The initial candidates are detected as fixed points from
the filter center frequency to the output instantaneous frequency.
The extractor selects the initial four candidates based on the SNR.
These are the candidates displayed in the proposed tools.

\end{document}